\documentclass[12pt]{iopart}
\usepackage{amsfonts}
\usepackage[normalem]{ulem}
\usepackage{bm}
\usepackage{graphicx}
\usepackage{soul}
\usepackage{color}
\usepackage{booktabs}
\usepackage{adjustbox}
\usepackage{tabularx}

\newcommand{\ket}[1]{\vert#1\rangle}

\def\opone{\leavevmode\hbox{\small1\kern-3.8pt\normalsize1}}
\usepackage{color}

\usepackage{appendix}
\begin{document}

\title{Measurement-device-independent quantum key distribution coexisting with classical communication
}

\author{R. Valivarthi$^{1,2}$, P. Umesh$^{1,}$\footnote{Current address: QuTech, and Kavli Institute of Nanoscience, Delft Technical University, Delft, The Netherlands}, C. John$^{3}$, K.~A.~Owen$^{1}$, V.~B. Verma$^{5}$, S.~W. Nam$^{5}$, D. Oblak$^{1}$, Q. Zhou$^{1,6}$ and W. Tittel$^{1,7}$}

\address{$^{1}$ Department of Physics and Astronomy, and Institute for Quantum Science and Technology, University of Calgary, Calgary, T2N 1N4, Canada}
\address{$^{2}$ The Institute of Photonic Sciences (ICFO), 08860 Casteldefels, Barcelona, Spain}
\address{$^{3}$ Department of Electrical and Computer Engineering, University of Calgary, Calgary, AB, T2N 1N4, Canada}
\address{$^{4}$ Jet Propulsion Laboratory, California Institute of Technology, Pasadena, CA, 91109, USA}
\address{$^{5}$ National Institute of Standards and Technology, Boulder, CO, 80305, USA}
\address{$^{6}$ Institute of Fundamental and Frontier Science, and School of Optoelectronic Science and Engineering, University of Electronic Science and Technology of China
(UESTC), Chengdu, China} 
\address{$^{7}$ QuTech, and Kavli Institute of Nanoscience, Delft Technical University, Delft, The Netherlands
}

\eads{\mailto{w.tittel${@}$tudelft.nl}}

\begin{abstract}
 The possibility for quantum and classical communication to coexist on the same fibre is important for deployment and widespread adoption of quantum key distribution (QKD) and, more generally, a future quantum internet. While coexistence has been demonstrated for different QKD implementations, a comprehensive investigation for measurement-device independent (MDI) QKD -- a recently proposed QKD protocol that cannot be broken by quantum hacking that targets vulnerabilities of single-photon detectors -- is still missing. Here we experimentally demonstrate that MDI-QKD can operate simultaneously with at least five 10 Gbps bidirectional classical communication channels operating at around 1550 nm wavelength and over 40 km of spooled fibre, and we project communication rates in excess of 10 THz when moving the quantum channel from the third to the second telecommunication window. The similarity of MDI-QKD with quantum repeaters suggests that classical and generalised quantum  networks  can  co-exist  on  the  same  fibre  infrastructure.
\end{abstract}


\maketitle

\section{Introduction}
The prospect of building a quantum internet, which promises information-theoretic secure communication \cite{Gisin2002} as well as blind or networked quantum computing \cite{Fitzsimmons}, is generating a rapidly increasing amount of academic and corporate development efforts \cite{EUflagship}. To minimise operating costs and hence facilitate deployment, it is important to benefit as much as possible from existing infrastructure. Starting in 1995, this has encouraged many experiments with deployed telecommunication fibre \cite{Muller95,Tittel98,Valivarthi2016}, and, since 1997, demonstrations of quantum key distribution (QKD) -- the most mature application of quantum networks -- together with classical data on the same fibre \cite{Townsend1997,Eraerds2010,patel2014, frohlich2015,wang2017,eriksson2018coexistence}. Yet, to date, comprehensive studies of the latter have been limited to so-called prepare-and-measure (P\&M) QKD \cite{Gisin2002}, in which one user,  Alice,  encodes a random string of classical bits into non-orthogonal quantum states of  photons, and the other  user,  Bob, makes  projection  measurements onto a set of randomly chosen bases.  Mapping measurement outcomes onto  bit  values  leads  to  the  so-called  raw-key---two  partially  correlated  sequences  of zeros and ones (one at Alice, and one at Bob)---and, after key distillation,  either to the creation of an error-free secret key, or to abortion of the key generation session. 

While the security of properly implemented P\&M QKD can be proven, it is threatened by blinding attacks - quantum hacking that exploits vulnerabilities of single-photon detectors to change their functioning \cite{Lydersen2010} (see \ref{app:protocol} for more information). This problem can be overcome by measurement-device-independent (MDI) QKD \cite{Lo2012MDI}, in which Alice and Bob both send photons to a central station, Charlie, who projects their joint state onto one or more of the four maximally entangled Bell states 
\begin{eqnarray}
\ket{\psi^\pm}&=&(\ket{01}\pm\ket{10})/\sqrt{2}, \\ 
\ket{\phi^\pm}&=&(\ket{00}\pm\ket{11})/\sqrt{2}. 
\end{eqnarray}
Here $\ket{0}$ and $\ket{1}$ denote two orthogonal quantum states, e.g. orthogonal polarisation or temporal modes. As in the case of P\&M QKD (or entanglement-based QKD\cite{Gisin2002}), any eavesdropping during photon transmission will lead to errors and shortening of the secret key -- possibly to zero length. However, beyond what is offered by all QKD protocols, this feature also holds in MDI-QKD if the actual measurement devices---that is the detectors---deviate from the ideal, including due to blinding attacks by Eve.

The proposal of the MDI-QKD protocol in 2012 triggered rapid experimental progress. The first proof-of-principle demonstrations were reported only a year later \cite{Rubenok2013,Liu2013,daSilva2013}, and the performance of MDI-QKD systems---including maximum distance, secret key rates, and robustness---has improved ever since \cite{Valivarthi2015,Yin2016,comandar2016quantum,tang2016measurement}. However, unlike for P\&M QKD, coexistence of MDI-QKD with
classical data on the same fibre has not yet been investigated in a comprehensive manner, neither experimentally nor through simulations. 
(But we note that spectraly multiplexed light was used in one MDI-QKD implementation to assess and compensate for polarisation transformations in the quantum channel, as well as to transmit a 1 MHz clock signal \cite{daSilva2013}.)

The difficulty of combining classical and quantum communication over the same fibre lies in the generation of noise photons that may mask the quantum data. While adequate spectral filtering can efficiently prevent all light at the classical data wavelength, assumed to be different from that used for the quantum channel, from reaching the single-photon detectors, this is no longer true for photons that are created by processes that convert the classical light to wavelengths within the quantum channel. Of particular concern is the interaction with phonons, so-called Raman scattering, which leads to the generation of noise over a wide range of wavelengths.
The scattered power in case of co- and counter-propagating classical and quantum channels, $P_{co}$ and $P_{ct}$, respectively, is given by \cite{frohlich2015}   
\begin{eqnarray}
P_{co} &=& P_{l}\beta\Delta \lambda \frac{(e^{-\alpha_Q L}-e^{-\alpha_{C} L})}{\alpha_{C} - \alpha_Q}\\
P_{ct} &=& P_{l}\beta\Delta \lambda \frac{(1 - e^{-(\alpha_{c}+\alpha_{Q}) L})}{\alpha_{C} + \alpha_Q}
\end{eqnarray}
\noindent
where $L$ is the fibre length,  $P_{l}$ is the average power launched in the classical channel, $\beta$ is the Raman scattering coefficient ($\beta$ depends on the wavelengths of the quantum and the classical channels as well as properties of the optical fibre), $\Delta \lambda$ is the bandwidth of the quantum channel, and $\alpha_Q$ and $\alpha_{C}$ are the fiber attenuation coefficients for quantum and classical channels, respectively. The photon scattering rate, $n$, and the scattered power, $P$, are related by $n hc/\lambda=P$, where $h$ is Planck's constant, $c$ the speed of light, and $\lambda$ the photon wavelength.
For bidirectional communication, allowing the exchange of classical data between Alice and Bob 
over a single fibre, 
the rates for co- and counter-propagating data have to be added: $P_{bi}=P_{co}+P_{ct}$. 

In this paper we experimentally demonstrate that measurement-device independent (MDI) QKD can operate simultaneously with at least five 10 Gbps bidirectional classical communication channels at around 1550 nm wavelength over 40 km of spooled fibre, and we project communication rates in excess of 10 THz when moving the quantum channel from the third to the second telecommunication window. As MDI-QKD is ideally suited for building cost-effective QKD networks with star-type topology, and can be upgraded into quantum-repeater-based networks \cite{repeaters}, our demonstration is a first step towards a future quantum network in which secret keys, or qubits, can be distributed over arbitrarily long distances, and using which networked quantum information processing and blind quantum computing will become possible.

\begin{figure*}[bbbb!]
\includegraphics[width=0.95\textwidth]{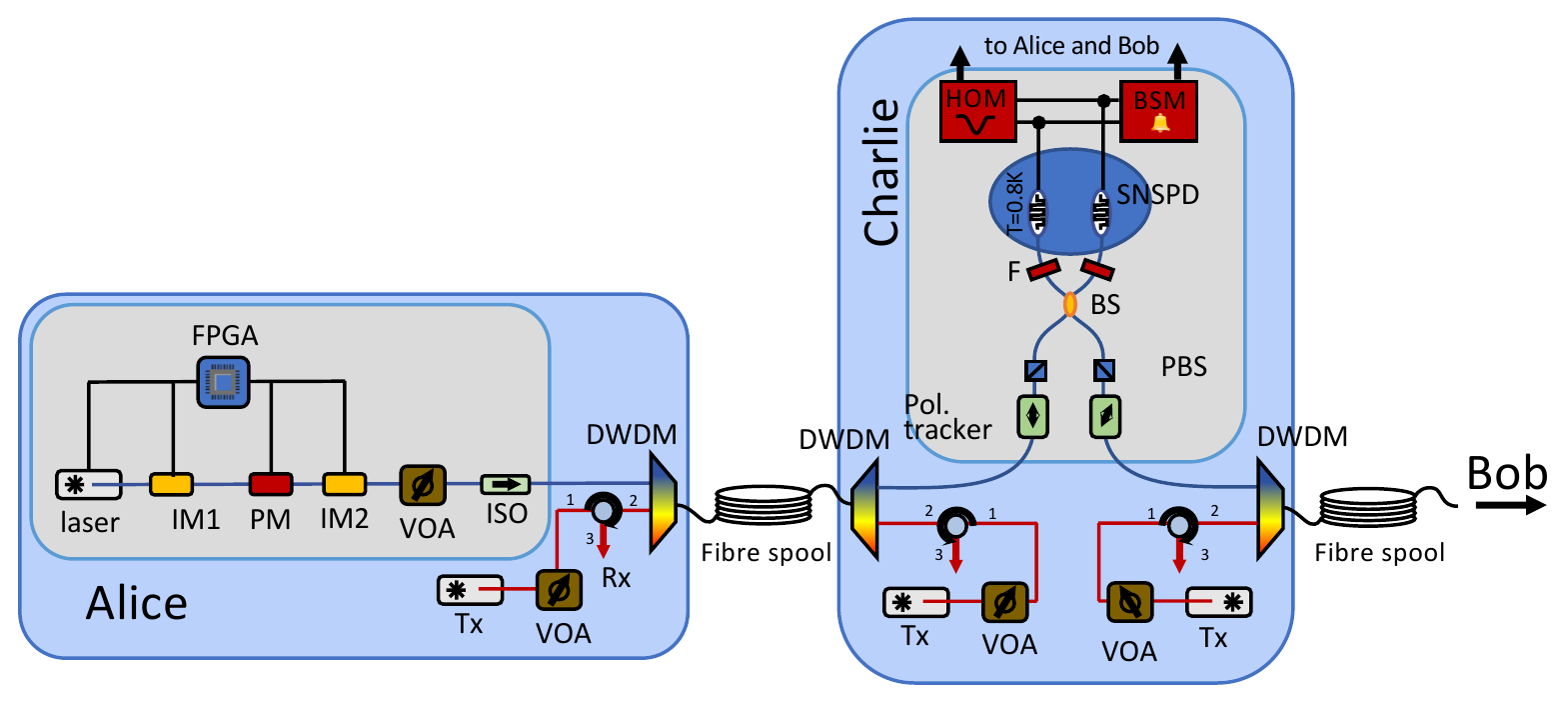}
\caption{\textbf{Experimental setup.} Only one sender unit and the central receiver is shown. See \ref{app:system} for details. Intensity modulator (IM), phase modulator (PM), variable optical attenuator (VOA), optical isolator (ISO), field-programmable gate array (FPGA), dense wavelength demultiplexer (DWDM), classical transmitter (Tx), classical receiver (Rx), polarizing beam splitter (PBS), beam splitter (BS), narrow spectral filter (F), superconducting nanowire single-photon detectors (SNSPD), Hong-Ou-Mandel dip measurement (HOM), Bell-state measurement (BSM).}
\label{fig:setup}
\end{figure*}

\section{Methods\label{methods}}
  
Our demonstration of coexistence with classical data is based on the MDI-QKD setup depicted in figure~\ref{fig:setup} and further detailed in \ref{app:system} (see also \cite{Valivarthi2017}). Additional classical communication channels are prepared using four 1548 nm DFB lasers, sending continuous-wave light from Alice to Charlie, from Charlie to Alice, from Bob to Charlie, and from Charlie to Bob. 
The launch power of each laser is chosen such that at the remaining power at the receiver side is an integer multiple of 2 $\mu$W -- the minimum power needed for a 10 Gbps link \cite{patel2014}. For instance, 10 $\mu$W at the receiver side corresponds to either one 50 Gbps channel, or to five 10 Gbps channels realized using different frequencies within the ITU grid. Provided neighbouring channels are chosen, the Raman noise created by all classical channels at the quantum channel wavelength 16 nm away can be considered equal, and it does therefore not matter over how many channels classical data is distributed.
Quantum and classical data are combined and split using dense wavelength division multiplexer (DWDM). 

\subsection{Raman noise}
\begin{figure*}[bbbb!]
\includegraphics[width=15.75cm, height=3.5cm]{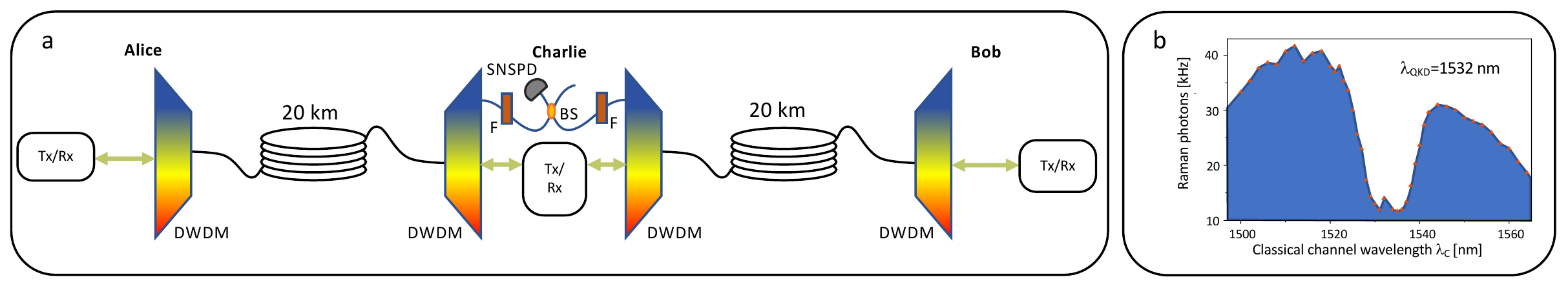}
\caption{\textbf{Crosstalk.} \textbf{a,} Schematics of the setup used for assessing crosstalk due to Raman scattering. Classical transmitter (Tx) and receiver (Rx), dense wavelength demultiplexers (DWDM), beam splitter (BS), narrow spectral filter (F), superconducting nanowire single-photon detector (SNSPD). Classical light was injected bi-directionally into two 20 km-long fibres (Corning SMF-28 standard telecommunication fibre) connecting Alice, and Bob, to Charlie. The launch power for each classical channel,  $\sim$8$\mu$W, was chosen so that each output power was 2~$\mu$W -- sufficient for 10 Gbps classical communication \cite{patel2014}. \textbf{b,} Raman noise measured using a single-photon detector at Charlie in a 6 GHz broad spectral channel centred at $\lambda_{QKD}$=1532.68 nm wavelength for different classical channel wavelengths $\lambda_{C}$.
}
\label{fig:crosstalk}
\end{figure*}

To assess the effect of Raman scattering on MDI-QKD, we first measured the noise in a narrow spectral window centred at 1532 nm---the operating wavelength of our MDI-QKD system---caused by strong light of various wavelengths propagating bi-directionally through 20 km-long standard telecommunication fibre between Alice and Charlie, and Bob and Charlie. The measurement is described in more detail in figure~\ref{fig:crosstalk}a.

\subsection{Experimental secret key rates} 
Next, we ran our QKD system over two different lengths of spooled fibre -- 2~$\times$~20~km, and 2~$\times$~40~km. As in the case of assessing cross-talk, the quantum channels between Alice and Charlie, and Bob and Charlie, were combined with pairs of bi-directional classical data channels. To test the worst case in which Raman noise is maximized, we used 1548 nm laser light for the data channel (this choice is motivated by the result of the measurement shown in figure \ref{fig:crosstalk}b), and to emulate different numbers of classical channels, we changed the power at each input in integer multiples of $\sim$ 8 $\mu$W ($\sim$ 20 $\mu$W), corresponding to 2~$\mu$W steps in  output power after 20 km (40 km) transmission. As shown before \cite{patel2014}, 2~$\mu$W suffices to operate one 10 Gbps data channel with bit error rates $\leq$10$^{-12}$, and having hence N time that power at the four receivers hence allows for N bi-directional 10 Gbps links between Alice and Bob.

For each configuration of fibre length and number of bi-directional 10 Gbps channels, emulated using continuous-wave light with appropriately chosen power, we created sifted keys and evaluated the secret key rate according to
\begin{equation}
R_{\inf}\geq [Q_{11}^Z [1-h_2(e_{11}^X)]-Q_{\mu \sigma}^Z f h_2(e_{\mu \sigma}^Z)].
\label{key rate equation}
\end{equation}
Here, $Q_{11}$ is the gain (the probability of a projection onto a Bell state) per emitted pair of qubits; $e_{11}$ the associated error rate; and the superscript indicates the jointly used basis (the Z basis features eigenvectors $\ket{0}$ and $\ket{1}$), and the X-basis eigenvectors ($\ket{0} \pm \ket{1})/\sqrt{2}$). Furthermore, $h_2(x)=-x~log_2(x)-(1-x)~log_2(1-x)$ is the binary Shannon entropy function; f = 1.14 is the efficiency of error correction; and the subscript ``inf'' denotes the assumption of infinitely long keys.

\subsection{Simulations}
We simulated secret key rates in the presence of classical communications using the code described in detail in our previous studies\cite{Chan2014, Valivarthi2017}. Noise caused by Raman scattering is taken into account by increasing the detector noise according to the results shown in figure~\ref{fig:crosstalk}. For simulations that require Raman noise within a quantum channel centered around 1310 nm wavelength and a classical channel within the C-band, we used experimental data published elsewhere \cite{wang2017}.

\section{Results}
\subsection{Raman noise}
The results of the measurements of the Raman noise are shown in figure~\ref{fig:crosstalk}b (the numerical data is listed in \ref{app:data}). Confirming previous observations \cite{frohlich2015}, we find Raman noise even if the quantum and classical channels are separated by many tens' of nanometers, and that the gain of the underpining interaction is reduced if the channel spacing is less than a few nanometers. Limiting classical channels to the extensively used C-band (extending from 1530 to 1565 nm wavelength), we furthermore see that the most cross-talk happens at a wavelength of approximately 1548 nm.

\subsection{Experimental key rates}
 
\begin{center}
\begin{figure}[h!]
\includegraphics[width=10cm]{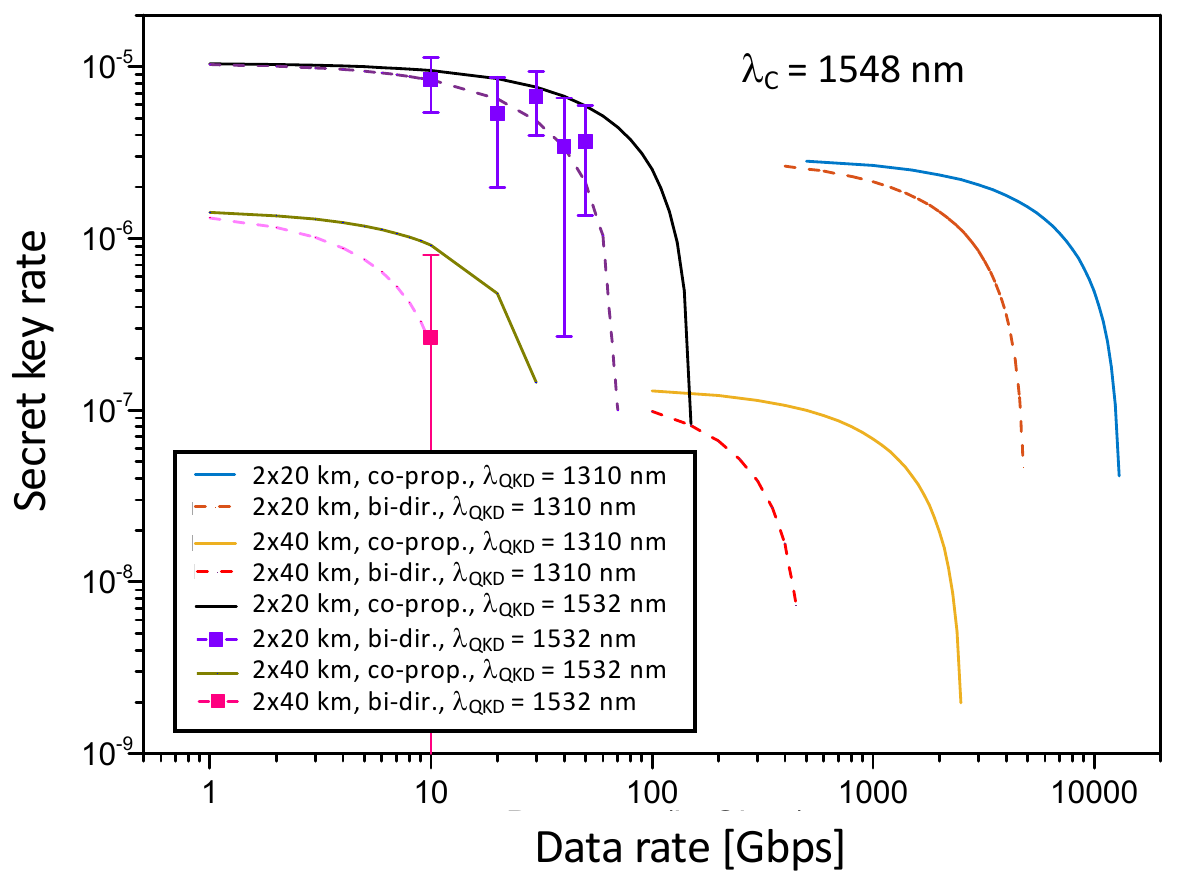}
\caption{\textbf{Results.} Predicted (lines) and experimentally obtained (squared) secret key rates (per clock cycle and assuming the infinite key limit) for different fibre lengths, data rates, wavelengths of the quantum channel, and assuming bi-directional or uni-directional (co-propagating) classical communication between Alice and Bob (connected via Charlie). Classical data is assumed to be at $\lambda_C$=1548 nm wavelength. Experimental error bars indicate one standard deviation and are obtained assuming Poisson detection statistics.
}
\label{fig:results}
\end{figure}
\end{center}

Secret key rates in the infinite (key length) limit, together with predictions based on an independent characterisation of the complete setup (no fits) are depicted in figure~\ref{fig:results} (the numerical data are listed in \ref{app:data}). 

\section{Discussion}
Most importantly, we find that MDI-QKD and bi-directional classical communication is possible over the same fibre. More precisely, we experimentally demonstrated positive secret key rates over a total of 40 km fibre together with the possibility for up to 50 Gbps bi-directional classical communication, and theoretically predicted positive secret key rates with up to 70 Gbps of classical data over the same fibre length. In addition, we demonstrated the possibility for QKD over a total of 80 km fibre with 10 Gbps of classical data. This is comparable to results obtained for P\&M QKD, e.g. in \cite{patel2014} where the possibility for secure key exchange over 70 km fibre distance and coexisting with 10 Gbps of bi-directional classical communication was demonstrated. However, the quantum-classical channel spacing was only of 2.5 nm in this case, resulting in approximately three times less Raman noise as compared to the worst-case scenario of 16 nm spacing chosen in our implementation (see figure~\ref{fig:crosstalk}). The apparent increased resilience of MDI-QKD to Raman noise may be due to the need for detecting two photons per key bit. However, the flip-side is a reduced key rate, at least as long as single-photon detectors with quantum efficiencies significantly below unity are employed.

The classical communication rate or, alternatively, the number of classical data channels at neighbouring spectral channels can straightforwardly be increased by 50\% by moving the classical data within the C-band from 1548 nm to 1565 nm wavelength, where Raman noise is reduced (see figure~\ref{fig:crosstalk}b). Furthermore, as shown by the simulations depicted in figure~\ref{fig:results}, the maximum classical data rate would increase by almost two orders of magnitude, e.g. for a total distance of 40~km from around 70 Gbps to around 5 Tbps, when shifting the QKD wavelength to 1310 nm wavelength (the classical data is assumed to be at 1548 nm wavelength, but changes within the C-band barely affect performance). In this configuration, increased photon transmission loss---normally degrading QKD performance---is more than compensated for by a reduction of Raman scattering.

Even better performance is expected when moving from bi-directional transmission of classical data to uni-directional transmission, where data co-propagates with QKD photons. In this case, most Raman photons, created in the region of highest laser power, i.e. close to Alice or Bob, would be absorbed in the fibre before arriving at Charlie's detector. As shown in figure~\ref{fig:results}---and still assuming a QKD wavelength of 1310 nm and classical data to be encode in the C-band---this would allow the distribution of secret keys together with classical communications over a total of 40 km  at more than 10 Tbps rate. This suffices for most applications.

We note that our QKD system currently employs pairs of fibres -- one fibre for clock synchronisation and announcement of successful measurements at Charlie, and one for quantum communication. However, our results shows that quantum and classical signals can be multiplexed into the same fibre. We also point out that the calculation of the secret key rate in Eq. \ref{key rate equation} assumes the limit of an infinitely long sifted key. This is in reality impossible, and an additional reduction that depends on the key length before post processing has to be taken into account \cite{Xu2014}. For instance, with $\sim$0.2 kbps of sifted key, as in our current setup over 2x20 km fibre, it would take $\sim$ 139 hours to pass the threshold between no secret key and secret key. While feasible, this is is impractical. The time can be reduced by two orders of magnitude by increasing the clock rate from its current value of 20 MHz to a few GHz. Current bottlenecks to this solution are the maximum clock rate of the (sequentially-operated) FPGAs in the QKD senders; limited accuracy (e.g. ringing) of the signals used to drive intensity and phase modulators; and the recovery time of the superconducting nanowire single-photon detectors. They can be overcome by more advanced FPGA programming, better electronics, and the use of detector arrays\cite{allman2015near}.

\section{Conclusion}
Our investigation establishes the possibility for MDI-QKD to coexist with classical communication on the same fibre. Moreover, as MDI-QKD shares an essential feature with quantum repeater-based communication -- the need for a Bell state measurement with photons that are created far apart -- it also shows that classical and generalised quantum networks can co-exist on the same fibre infrastructure. We additionally note that MDI-QKD is ideally suited for building QKD networks with star-type topology in which several users are connected to the same central measurement node (Charlie). Using optical switches, it becomes then possible to connect any pair of users on demand. As users only need sender modules but no receivers (the latter will be located at the central node and be accessible to all users), this solution is both simpler and more cost-effective than the creation of a fully connected network using P\&M QKD, which requires all users to have both a sender and a receiver module. Hence, our demonstration increases the commercial viability of MDI-QKD and, more generally, quantum communications, will facilitate the adoption of the new quantum technology, and therefore constitutes an important step towards a world in which quantum information processing will help meeting challenges in secure data transmission, and will provide opportunities for unparalleled data processing. 
 
\section*{Acknowledgements}
The authors thank Vladimir Kiselyov for technical help, and Alberta Innovates Technology Futures (AITF), the National Science and Engineering Research Council of Canada (NSERC), and the Netherlands Organization for Scientific Research (NWO) for financial support. Furthermore, WT acknowledges funding as a Senior Fellow of the Canadian Institute for Advanced Research (CIFAR), and VBV and SWN acknowledge partial funding for detector development from the Defense Advanced Research Projects Agency (DARPA) Information in a Photon (InPho) program. Part of the research was carried out at the Jet Propulsion Laboratory, California Institute of Technology, under a contract with the National Aeronautics and Space Administration (NASA).


\appendix
\section{Quantum hacking}\label{app:protocol}


During recent years, it became apparent that the ideal security of QKD protocols may not carry over to actual implementations. By exploiting deviations---possibly induced by the eavesdropper herself---of the used technology compared to the assumptions that underpin the security proof, it may be possible for Eve to acquire full information about the key without abortion of the QKD session.  In particular, single-photon detectors have been shown vulnerable to such side-channel attacks \cite{Lydersen2010}, and while countermeasures to certain attacks have been demonstrated \cite{yuan2010avoiding}, their effectiveness remains questionable \cite{Lydersen2010b}. Furthermore, a countermeasure can only be implemented once an attack has been discovered, leaving all keys exchanged in the intermediate period insecure. In light of this, alternative protocols whose security against side-channel attacks is rooted in fundamental quantum mechanical principles have been proposed. Currently the most practical of these protocols is measurement-device-independent (MDI) QKD\cite{Lo2012MDI}, which is described next.

 
\section{The MDI-QKD system}\label{app:system}
\subsection{Qubit preparation at Alice and Bob}
By shaping (using intensity and phase modulators, IM, PM) and attenuating (using a variable optical attenuator, VOA) phase-randomized pulsed of light emitted by laser diodes driven from below threshold, Alice and Bob randomly create qubit states encoded into superpositions of early and late temporal modes:
\[
\ket{0}\equiv\ket{e};\:
\ket{1}\equiv\ket{\ell}
\]
\[
\ket{+}\equiv\big (\ket{e}+\ket{\ell}\big )/\sqrt{2}; \:
\ket{-}\equiv\big (\ket{e}-\ket{\ell}\big )/\sqrt{2}.\\
\]
The first two states are eigenstates of the Z-basis, and the two latter of the X-basis. Each temporal mode extends over 500 ps, and early and late modes are separated by 2.5 ns. 
Security against photon number splitting attacks \cite{Gisin2002}, exploiting that more than one photon may may be present in an attenuated laser pulse, is derived by randomly changing their mean photon numbers between three different values, according to the three-intensity decoy-state protocol described in \cite{Xu2014}. Furthermore, optical isolators (ISO) that prevent light from the outside to enter the QKD senders protect against Trojan horse attacks \cite{Gisin2002}. 

Random numbers, both for selecting qubit states as well as intensity levels of attenuated laser pulses, are created off-line using a quantum random number generator \cite{zhou2017practical}, stored in field programmable gate arrays (FPGA) within each sender module, and then used to determine phase and intensity modulator settings.

\subsection{Qubit measurement at Charlie}
A projection onto the $\ket{\psi^-}=(\ket{e\ell}-\ket{\ell e})/\sqrt{2}$ Bell state takes place if two photons, one from Alice and one from Bob, are detected behind a 50/50 beamsplitter (BS) in different temporal modes -- one early, and one late. To ensure the required indistinuishability of the two photons, we employ several automated feedback loops. First, arrival time differences are measured using Hong-Ou-Mandel (HOM) interference, and synchronization is maintained by delaying the clock signal sent from Charlie to either Alice or Bob. Furthermore, polarization indistinguishability is ensured by means of polarizing beam-splitters (PBS, with feedback to maximize transmission) and polarization-maintaining fibres that connect to the 50/50 beam-splitter that is at the heart of the Bell-state measurement. In addition, we verify the frequency difference between Alice's and Bob's temperature-stabilized laser diodes every 5 minutes and, if necessary, reduce it to less than 10 MHz. 

Photons are detected using WSi superconducting nanowire single-photon detectors (SNSPDs) cooled to 0.8 K in a sorption cooler \cite{marsili2013detecting}. They feature system efficiencies of around 50\%, dark counts of around 100~Hz, and detection time jitter of 100~ps. 
Successful Bell-state measurements are communicated to Alice and Bob using laser pulses sent over additional fibre.

\subsection{Key sifting}
The first step in key sifting is the reduction of the local bit strings at Alice's and Bob's to those that describe the states of photons that were detected in Charlie's Bell-state measurement. In order to avoid  memory-intensive storage of time-tagged data that characterizes all photon states--most of which will be discarded during this step--Alice and Bob send the information of their prepared qubits (with the exception of time) into first-in-first-out (FIFO) buffers in their FPGAs while the corresponding qubits are sent to Charlie. The delays in the buffers equal the combined time required by the qubits to reach Charlie, and by the BSM signals to travel back to Alice or Bob. A simple logic operation then allows singling out only qubit generations that resulted in a successful BSM -- only those are further processed during subsequent basis reconciliation. 

\section{Data}\label{app:data}

\begin{table}[hhhh!]
\begin{center}\caption{Raman noise measured at Charlie in a 6 GHz wide  spectral window centered at 1532 nm wavelength for different classical channel wavelengths.}
\begin{tabular}{@{}ll||ll}
\br
Wavelength [nm]            & Noise counts [kHz] & Wavelength [nm]            & Noise counts [kHz]\\ \mr 
1500 & 33.33        & 
1505 & 35.33        \\ 
1510 & 40.67        &
1515 & 41.33        \\ 
1520 & 38.00        & 
1525 & 30.00        \\ 
1530 & 13.00        & 
1535 & 11.67        \\ 
1540 & 23.67        & 
1545 & 31.00        \\ 
1550 & 28.67        & 
1555 & 26.33        \\ 
1560 & 23.00        & 
1565 & 17.67        \\ 
\br
\end{tabular}
\end{center}
\end{table}
\normalsize

\begin{table}[hhhh!]
\caption{Experimentally obtained secret key rate ($R_\infty$) with number of co-existing 10 Gbps channels, N, for different transmission lengths of spooled fibre.}
\begin{tabular}{@{}lll}
\br
N & 2 $\times$ 20 km        & 2 $\times$ 40 km          \\ \mr
0 & 1.13E-05 $\pm$ 5.52E-06 & 1.72E-06 $\pm$ 6.16E-07 \\ 
1 & 8.37E-06 $\pm$ 2.93E-06  & 2.66E-07 $\pm$ 5.35E-07 \\ 
2 & 5.34E-06 $\pm$ 3.36E-06  &                          \\ 
3 & 6.66E-06 $\pm$ 2.69E-06  &                          \\ 
4 & 3.43E-06 $\pm$ 3.16E-06  &                          \\ 
5 & 3.66E-06 $\pm$ 2.29E-06  &                          \\ 
\br
\end{tabular}
\end{table}

\newpage
\section*{References}
\bibliography{coexisting_mdiqkd_ref}



\end{document}